\documentclass[aps,prb,showpacs,twocolumn]{revtex4}
\usepackage{graphicx}
\usepackage{amsmath}
\usepackage{amsfonts}
\usepackage{color}

\begin{document}
\title{Topological surface states scattering in Antimony}

\author{Awadhesh Narayan, Ivan Rungger, and Stefano Sanvito}
\affiliation{School of Physics and CRANN, Trinity College, Dublin 2, Ireland}

\date{\today}

\begin{abstract}
In this work we study the topologically protected states of the Sb(111) surface by using {\it ab-initio} transport theory. 
In the presence of a strong surface perturbation we obtain standing-wave states resulting from the superposition 
of spin-polarized surface states. By Fourier analysis, we identify the underlying two dimensional scattering processes and 
the spin texture. We find evidence of resonant transmission across surface barriers at quantum well states energies 
and evaluate their lifetimes. Our results are in excellent agreement with experimental findings. We also show that 
despite the presence of a step edge along a different high symmetry direction, not yet probed experimentally, the surface 
states exhibit unperturbed transmission around the Fermi energy for states with near to normal incidence.   
\end{abstract}

\maketitle

{\it Introduction.---}In the past few years topological insulators (TIs) have become an intensely studied field of condensed matter 
physics.\cite{review-kane,review-zhang} The unique metallic states at the surface of a TI can be used both as a 
table-top playground to prove fundamental concepts developed for particle physics and also as new materials platform 
for intriguing quantum applications in both spin-electronics and quantum computing.\cite{kane-majorana, zhang-monopole, spin-pesin} The first three-dimensional TI to be experimentally realized was a 
Bi-Sb alloy \cite{hsieh08}, following a theoretical prediction by Fu and Kane.\cite{fukane-inv} The topological nature of the 
alloy is inherited from the parent element antimony, which has a non-trivial principal topological invariant $\nu_0$, 
whereas bismuth has a trivial $Z_2$ invariant.\cite{bisb-kane} 

Although Sb itself is a semimetal, its $(111)$ surface hosts two spin-polarized bands, which extend around the Fermi 
energy, $E_\mathrm{F}$. These form a single distorted Dirac cone where the lower bands are lifted upwards. 
The electronic properties of Sb(111) thin films have been investigated both theoretically and 
experimentally.\cite{sb111-wang, sb111-liu, sb111-sasaki, sb111-chiang} In a recent experiment, Seo {\it et al.} 
demonstrated that the topological surface states are extraordinarily insensitive to the presence of surface 
barriers.\cite{sb-yazdani} They probed the extended nature of Sb$(111)$ surface states by using a scanning tunneling
microscope (STM) and found that these transmit across surface atomic steps with a high probability. Their analysis 
of the standing-wave states on surface terraces revealed the novel chiral spin texture of the two surface states, 
consistent with earlier Angle-resolved photoemission spectroscopy (ARPES) measurements.\cite{hsieh09} 

In this Rapid Communication we seek to theoretically recreate the above mentioned experiment by using {\it ab-initio} transport 
theory, and show that we can reproduce the formation of quantum well states and their life-times, as well as the 
wave lengths and phase shifts of the scattering states. Thereby we demonstrate that by first principles calculations 
one can describe the correct scattering properties of such topologically protected surface states. In addition of comparing 
our results favorably to the experiments, we predict the scattering properties of these states in presence of a surface 
perturbation along a direction orthogonal to the one probed experimentally.

{\it Computational methods.---}Our first-principles electronic structure calculations are performed with density functional theory (DFT) 
using the local density approximation (LDA) to the exchange-correlation functional. We employed the {\sc siesta}
package, which implements a linear combination of atomic orbitals basis set.\cite{soler-siesta} Spin-orbit 
interaction, essential to describe the surface states, has been included via the on-site approximation.\cite{sanvito-soc} 
We include Sb $5s$ and $5p$ as the valence elecrons. In the slab geometry, there is a 10 \AA{} vacuum in the supercell, 
to avoid interaction between periodic images. The transport properties are then calculated by using {\sc smeagol}, which 
combines the non-equilibrium Green's function (NEGF) method with DFT.\cite{sanvito-smeagol1, sanvito-smeagol2, sanvito-smeagol3} 
In {\sc smeagol} the scattering region is attached to one or more semi-infinite electrodes via self-energies. The charge density is 
calculated by integrating the non-equilibrium Green's function, along a contour in the complex energy plane. For this we use, 16 
energy points on the complex semicircle, 16 points along the line parallel to the real axis and 16 poles. Periodic boundary conditions 
are employed in directions orthogonal to the transport direction, while using open boundary conditions along transport direction allows us to simulate single scatterers.\cite{sanvito-Si-on} We use an equivalent temperature of 300 K for broadening in the Fermi 
distribution. Our order-$N$ implementation of the code allows us to treat large systems.\cite{sanvito-Si-on} We use a double-$\zeta$ 
polarized (DZP) basis set, with a cutoff energy of $300$ Ry for the real space mesh. We have carefully checked convergence 
of our results with respect to all the parameters used. 

\begin{figure}[h]
\begin{center}
  \includegraphics[scale=0.40]{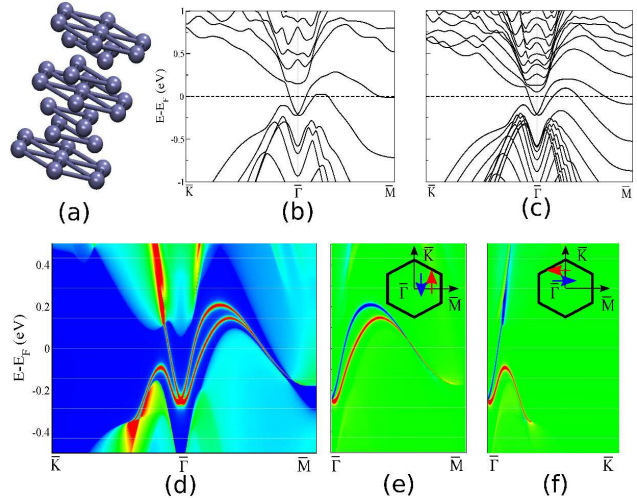}
  \caption{(Color online) (a) Structure of antimony in the hexagonal setting. The atoms form a bilayer structure with the intralayer 
  distance as $1.51$ \AA{} and the interlayer distance as $2.25$ \AA{}. Band structure for (b) six bilayers and (c) twelve bilayers thick 
  slabs along  $\bar{K}-\bar{\Gamma}-\bar{M}$ directions. (d) Simulated ARPES from a semi-infinite slab. The distorted Dirac cone is 
  found to be comprised of spin-polarized surface bands. Here and henceforth warmer colors represent higher PDOS (red represents 
  largest values, blue lowest ones, with the color scale in between being linear). Spin-resolved ARPES along (e) $\bar{\Gamma}-\bar{M}$ 
  and (f) $\bar{\Gamma}-\bar{K}$ directions showing the opposite spins of the two surface bands along the directions indicated by arrows 
  in the inset of the figures. In this case, red and blue colors indicate up and down spins, respectively.} \label{prelim}
\end{center}
\end{figure}

Sb crystallizes in a rhombohedral structure [space group $D^{5}_{3d}(R\bar{3}m)$] 
with two atoms per unit cell. An alternate way to represent its structure is in an hexagonal setting with the unit cell comprising six 
atoms. This representation is particularly useful to construct two-dimensional slabs, which are made of Sb bilayers, as shown in 
Fig.~\ref{prelim}(a). The inter-bilayer coupling is weak and it is possible to create surface steps which are a single bilayer 
high.\cite{sb-yazdani, sb-manoharan} The bulk structure was relaxed using the Vienna {\it ab-initio} simulation package 
({\sc vasp}) \cite{kresse-VASP}, until the forces were less than $0.01$ eV/\AA.   
 
{\it Results.---}We begin by calculating the surface band structure of six and twelve bilayers thick slabs of Sb [Figs.~\ref{prelim}(b) and (c)], 
by using a $10\times10$ in-plane $k$-point grid. The distorted Dirac cone at $\bar{\Gamma}$ is gapless indicating minimal interaction 
between the top and bottom surfaces of the slab. The surface band structure matches well previous {\it ab-initio} calculations.\cite{sb111-wang, sb111-liu, sb111-chiang} We find the Dirac point at an energy of about 210~meV below $E_\mathrm{F}$ for six and 
twelve bilayer slabs. In order to simulate the ARPES spectrum of an infinitely thick slab we perform a {\sc smeagol} calculation for the 6 
bilayer slab, where we attach semi-infinite Sb electrodes at the bottom layer via self-energies. The ARPES spectrum is then 
obtained by calculating the projected density of states on the surface atoms, and the result is shown in Fig.~\ref{prelim}(d).
The spin-resolved ARPES [Fig.~\ref{prelim}(e) and (f)] shows that the two surface bands carry opposite spin and exhibit the characteristic 
spin texture associated with topologically non-trivial materials. Furthermore, it can be seen that the surface states are more pronounced 
close to the $\bar{\Gamma}$ point. Once the two bands turn downwards from their maximum point, they are less localized on the surface 
due to their hybridization with the bulk bands. This matches with ARPES experiments, where at $\bar{M}$, no surface states are 
found.\cite{hsieh09} Thus, in both scattering and ARPES experiments, one would expect the dominant contributions to come from an 
area around the center of the BZ, with a radius of about one third of the length of BZ along $\bar{\Gamma}-\bar{M}$ direction. We note that 
the good agreement with the ARPES experiments shows that the LDA exchange-correlation functional is appropriate for this system.

\begin{figure}[h]
\begin{center}
  \includegraphics[scale=0.40]{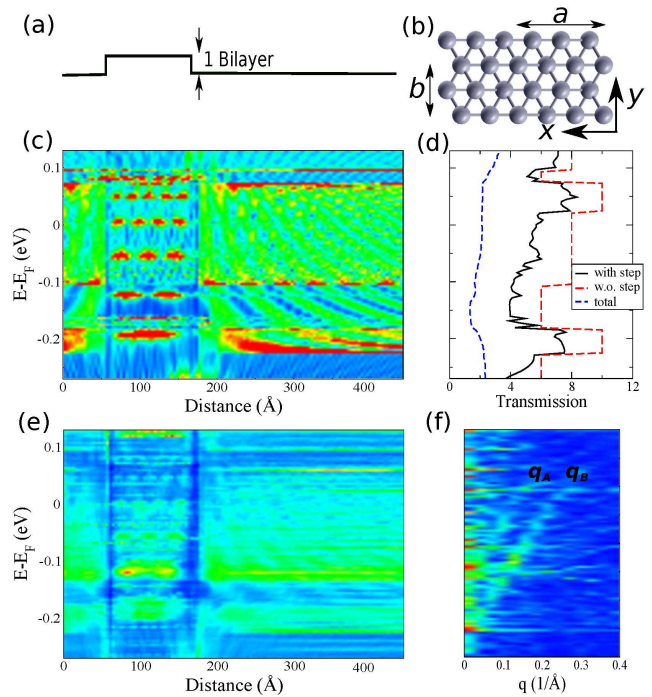}
  \caption{(Color online) (a) A schematic of the setup showing one bilayer high surface perturbation on a twelve bilayer slab. (b) Top view of the slab; surface step runs along $y$ for the first setup and along $x$ for the second. (c) PDOS for surface atoms with a single step adjacent to a flat region for the first setup at $k_{\bot}=0$. The quantization of the energy levels in the step is clearly seen, with accompanying phase shifts in the adjacent flat region. (d) Transmission at $k_{\bot}=0$ with and without the step indicating finite scattering due to the step. Average of transmission over all $k_\bot$-points is also shown. (e) PDOS for surface atoms averaged over all $k_{\bot}$. (f) Fourier transform of PDOS data in (e) over the flat region reveals the different allowed scattering wave vectors. The most prominent features, $q_A$ and $q_B$, are nearly linear with slopes equal to $1.1$ eV\AA{}.} \label{terrace-gamma-m}
\end{center}
\end{figure}

Next we simulate a step perturbation on the surface for two different directions: In the first case the step runs 
along the $y$-direction [see Fig.~\ref{terrace-gamma-m}(b)], so that the transport is along the $x$-direction, which is the 
same as in the experiment in Ref.[13]; and as a second orientation we choose the orthogonal direction (step running 
along $x$), to evaluate the effect of the orientation on the scattering. We have relaxed the step geometry 
for a smaller scattering region, but the atomic displacements were only minor, and henceforth we use the unrelaxed 
step configuration. A single bilayer high step is created over a length of $120$ \AA{}. The adjacent flat region 
extends over $270$ \AA{}. The setup consists of a $13$ bilayer-thick region with a short $12$ bilayer-thick region 
on the left and a longer one on the right, as shown schematically in Fig.~\ref{terrace-gamma-m}(a) (for the actual 
setup see Supplemental Material (SM) \cite{supplement}). We attach semi-infinite leads on left-hand and right-hand sides 
of the scattering region, by means of the self-energies calculated by {\sc smeagol}. This enables us to simulate isolated 
scatterers, thus avoiding the use of periodic boundary conditions along the transport direction.\cite{sanvito-Si-on} 

The total projected density of states (PDOS) $\mathcal{N}_{total}$ is obtained by integrating over all $k_{\bot}$-points perpendicular 
to the transport direction
\begin{equation}
\mathcal{N}_{total}(E)=\int_{k_{\bot}}{\mathcal{N}_{k_{\bot}}}(E)~dk_{\bot}.
\end{equation}
Analogously the total transmission is given by $T_\mathrm{total}(E)=\int_{k_{\bot}}T_{k_{\bot}}(E)dk_{\bot}$.
We note that in the first orientation of the step $k_\bot$ runs along the $\bar{\Gamma}-\bar{K}$ direction and the transport direction is parallel to $\bar{\Gamma}-\bar{M}$ in reciprocal space, whereas in the second orientation $k_\bot$ runs along $\bar{\Gamma}-\bar{M}$ and the transport direction is parallel to $\bar{\Gamma}-\bar{K}$. 
While we find that 3 $k_\bot$-points are sufficient for obtaining a converged self-consistent potential, we need many more $k_\bot$-points for accurately integrating $\mathcal{N}_{total}$ for a given potential, where we therefore use 200 $k_\bot$-points.

The $\mathcal{N}_{k_{\bot}}$ of the atoms on the top surface for $k_\bot=0$ is shown in Fig.~\ref{terrace-gamma-m}(c). 
The quantum well states formed by quantization of the energy levels in the step region are clearly visible, and 
extend over the energy window in which the two topological surface bands exist. The PDOS on the 
adjacent flat region shows oscillatory behavior typical of one-dimensional scattering barriers \cite{sanvito-Si-on}, and it has phase 
shifts at energies corresponding to the allowed energy states in the step region. At those energies we also find resonant transmission across quantum well states, visible as peaks in the transmission curve in Fig~\ref{terrace-gamma-m}(d). This matches the experimental 
observation of resonant tunneling through the surface barrier at those energies. It indicates a remarkably long phase coherence length of hundreds of angstroms, which is due to the extended 
nature of topological surface states.\cite{sb-yazdani} Further, the change of phase when the states are reflected from the barrier is nearly zero. 
Over the entire energy range there is clearly a rather large amount of scattering caused by the 
step and the transmission drops significantly below the value in absence of step.

We have verified that the single bilayer step at the top surface perturbs the bottom 
layer minimally and there is only small coupling between the two surfaces even in the presence of the surface step.\cite{supplement}

The integrated $\mathcal{N}_{total}$ is shown in Fig.~\ref{terrace-gamma-m}(e). The main features corresponding to the quantized energy levels 
can still be identified at almost the same energies found for $k_\bot=0$, but are broadened and less pronounced. The broadening of energy levels for $k_{\bot}=0$ is 8-12 meV, while for the total it increases
by a factor of about three. In experiment, these were found to lie between 20-45 meV.\cite{sb-yazdani} Hence, the states lifetimes, which 
are inversely proportional to their broadening, agree quantitatively with those found experimentally. This shows that the scattering properties of the step are well-reproduced in our 
calculations.
We note that quantitatively the results for a 12 bilayer slab differ somewhat from the ones one would obtain for a semi-infinite surface due to the different surface band structures [Fig. \ref{prelim}(c) vs. \ref{prelim}(d)]. Firstly, the linear dispersion region of the surface states around $\Gamma$ is more extended in energy for the semi-infinite surface, and this results in quantum well states spacing remaining constant, whereas for the 12 bilayer slab the spacing reduces with increasing energy. Moreover for 12 bilayers we find band edges at -185~meV and -108~meV, which are absent in the semi-infinite system, and which give strong contributions to the 12 bilayer PDOS.
Due to the enhanced PDOS at these band edges the phase shifts corresponding to resonant tunneling are not as 
clearly pronounced for $\mathcal{N}_{total}$ as for $k_{\bot}=0$.
Finally, whereas for the semi-infinite slab there are only two discrete surface bands around $E_F$, and all the other bands are diffuse, for the 12 bilayer slab clearly the number of distinct bands is larger, which as we will show leads to many more features in the scattering. These lead to discrete short-wavelength scattering processes which would be absent for a semi-infinite slab. 

In order to analyze the scattering mechanism and probe the chiral spin texture
of the surface bands, we evaluate the Fourier transform (FT) of
$\mathcal{N}_{total}$ along the flat region on the right side of the step,
and the result is shown in Fig.~\ref{terrace-gamma-m}(f).  For a given energy
and $k_\bot$ ($k_\bot$ is conserved during scattering) there are standing waves
with a wave vector $q$ for every pair of scattering states in the Sb electrodes
with opposite group velocities. For each such pair of states with indices $i$
and $j$ we get the norm of scattering vector $q_{ij}=\left|k_i-k_j\right|$, at which therefore the
FT has enhanced amplitude.  In Fig.~\ref{terrace-gamma-m}(f) we
find a rather large amount of noise at low $q$, partly due to the fact that we
integrate $\mathcal{N}_{total}$ over a finite number of $k_\bot$-points,
nevertheless we can identify two prominent features that are preserved for
increasing $q$. The first is an enhanced amplitude starting at about -210 meV
and then increasing linearly with $q$ with a slope of $1.1$ eV\AA, and the
second is an equivalent enhancement with the similar slope starting at -110
meV.  These two prominent scattering wave-vectors are also found in the
experiments of Seo {\it et al.}, where they are labeled $q_B$ and $q_A$,
respectively. $q_B$ is attributed to
scattering between the surface states close to $\bar{\Gamma}$ having the same spin,
but opposite momentum direction. In this case the scattering state
momenta are not equal, unlike in conventional scattering, because of the unique
spin texture of the surface states resulting in an asymmetric band structure for a given spin.
The other scattering wave-vector ($q_A$) is attributed to scattering between
neighboring hole pockets away from $\bar{\Gamma}$. The calculated slopes of $q_A$ and $q_B$ are remarkably close to the value of
$1.2$ eV\AA{} found experimentally.\cite{sb-yazdani} 
We note that to explain the origin of both $q_{A}$ and $q_{B}$, one needs to go beyond the one-dimensional model invoked by Seo {\it et al.} and a two-dimensional treatment is required.

\begin{figure}[h]
\begin{center}
  \includegraphics[scale=0.40]{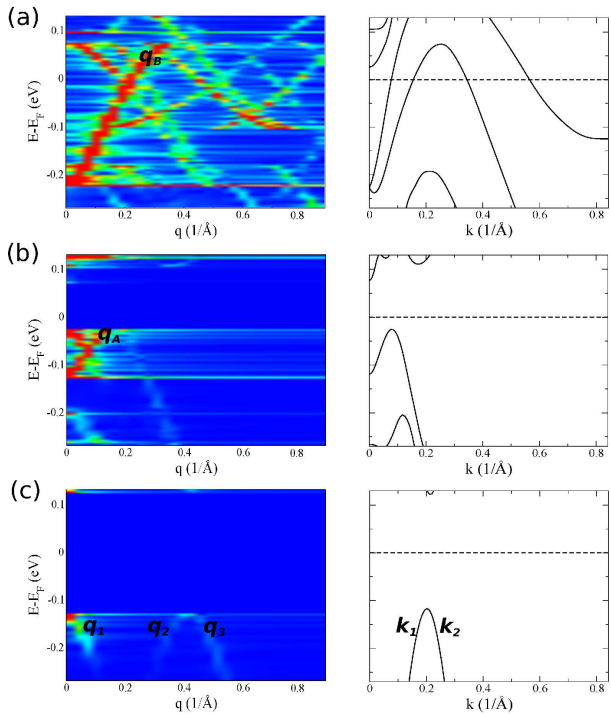}
  \caption{(Color online) The allowed scattering wave-vectors obtained from the Fourier transform of PDOS data at 
  (a) $k_{\bot}b=0$, (b) $k_{\bot}b=0.16\pi$ and (c) $k_{\bot}b=0.50\pi$. The panels on the right show the corresponding band structures at the same $k_\bot$
  for the 12 bilayer slab electrodes.} \label{ft-superpose}
\end{center}
\end{figure}

An STM experiment measures the total scattering in all the reciprocal space directions. In order to 
establish which $k_\bot$-points give rise to the $q_A$ and $q_B$ features in the average, we decompose the 
allowed scattering processes along different $k_\bot$ directions.
To illustrate the general scattering mechanism, in Fig. ~\ref{ft-superpose}(c) we show the Fourier transform for an arbitrary $k_\bot$-point, and the bands of the Sb electrodes along the transport direction for the same $k_\bot$. As an example, we consider the energy at -150 meV, where in the band structure we find 2 bands crossing for positive $k$, at $k_1$ and $k_2$. In the Fourier transform we find 3 scattering vectors: $q_1=k_2-k_1$, $q_2=2 k_1$ (obtained by $k_1$ scattering to $-k_1$), and similarly $q_3=2 k_2$. The amplitude is highest towards the band edge, where the PDOS is maximal. Analyzing the FT for all $k_\bot$-points, we find that most such features found for a single $k_\bot$ disappear when averaging, except for $q_A$ and $q_B$. This shows that the scattering processes visible in STM experiments are only a small subset of all processes occuring. Therefore no visible scattering from defects in an STM experiment does not imply perfect transmission, since the absence of standing wave patterns can also be due to the fact that the features may be broadened upon integration over $k_{\bot}$, even for substantial scattering for each single $k_\bot$.

There are two key factors, which decide which $q$ vectors dominate.
Firstly, for a given $k_\bot$ there needs to be a high scattering between the initial ($k_i$) and final ($k_j$) states. This is the case if the spins in the two states are aligned and if the surface-PDOS for them is large. This, therefore, excludes bulk states and states with opposite spins. Secondly, for the features to remain prominent when integrated over $k_{\perp}$, it is necessary that there is an extended region in the BZ where these features are found more or less unchanged. This is true when the change of band structure along $k_{\perp}$ is small, which is the case close to a maximum or a minimum, i.e. when $\partial E/\partial k_{\perp}=0$, and ideally when the band curvature is small ($\partial ^2 E/\partial  k_{\perp}^2$ is small).
We have verified that the above two conditions are indeed satisfied for $q_A$ and $q_B$.
From the 12 layer slab band structure along $\bar{\Gamma}-\bar{K}$ [Fig. \ref{prelim}(c)] we identify two such features: the first is the minimum at about $k_\bot=\bar{\Gamma}$, and the second is the maximum at $k_\bot b=0.16\pi$.
In Figs. ~\ref{ft-superpose}(a) and (b) we show the scattering wave vectors for these two $k_\bot$. We clearly identify the features leading to $q_A$ and $q_B$. The remaining features in the FT disappear under averaging, since for these the aforementioned conditions are not satisfied. We note that a related study has been recently performed by Takane and Imura by using a low-energy Dirac theory, where they find perfect transmission at all incidence angles for a hyperbolic step.\cite{takane-step} However, as noted by the authors, their analysis is valid in the long wave length regime, while we focus on atomic-scale terraces. Our results clearly demonstrate that there is scattering between states on the same Dirac cone for $k_\bot\ne0$, which leads to the appearance of the $q_{A}$ scattering vector. The fact that such a scattering vector is found prominently also in experiments indicates that for non-normal incidence the states are not perfectly transmitted, in agreement with our findings.

\begin{figure}[h]
\begin{center}
  \includegraphics[scale=0.40]{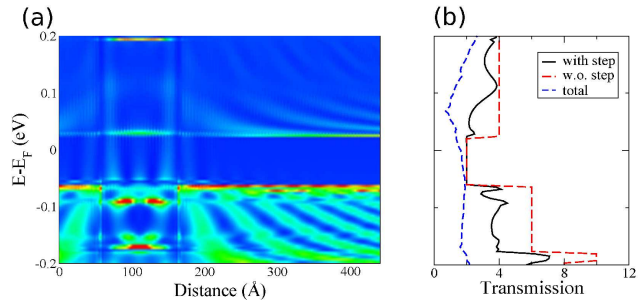}
  \caption{(Color online) (a) PDOS at $k_\bot=0$ for surface atoms with a single step adjacent to a flat region, with the step extending along the $x$ direction in Fig. \ref{terrace-gamma-m}(b). There is an energy region from -60 to 20~meV with no scattering and hence no standing wave states. (b) Transmission at $k_{\bot}=0$, indicating the perfect transmission around $E_F$ even in presence of a surface barrier, the defining feature of a topological state. Total transmission also shows minimal scattering in that energy range.} \label{terrace-gamma-k}
\end{center}
\end{figure}

Finally, we create a step on the Sb(111) surface along the $x$-direction. In this case the 
slab consists of six bilayers, since simulating a 12 bilayer scattering region along this direction is beyond our 
computational resources (see SM for a description of the setup \cite{supplement}). This direction is promising since 
there is an energy window [-60~meV to 20~meV in Fig.~\ref{prelim}(b)] over which only a single 
spin-polarized surface state exists for $k_{\bot}=0$ (note that in this case the transport direction is parallel to the $\bar{\Gamma}-\bar{K}$ line in reciprocal space), which is reminiscent of the prototypical TI Bi$_2$Se$_3$. The $\mathcal{N}_{k_{\bot}}$ of the surface atoms for $k_\bot=0$ is plotted in Fig.~\ref{terrace-gamma-k}(a). In the energy range of a single spin-polarized state there is 
no scattering, which is the hallmark of a topological surface state. From -170~meV to 70~meV quantum well 
states are found as a result of superposition between the two surface bands, with a mechanism analogous to that 
in the first step orientation. Fig.~\ref{terrace-gamma-k}(b) shows the transmission at $k_{\bot}=0$ with 
and without the surface step. Remarkably, there is a perfect transmission in this energy window, despite the presence 
of the strong surface perturbation in the form of an extended single bilayer high surface step. This can be explained by invoking 
the general principle that disorder which does not break time reversal symmetry can not localize a single 
topologically protected surface state. Away from this energy region, there is substantial scattering caused by the 
step and the transmission drops down from the value in its absence. The total transmission averaged 
over all $k_\bot$-points is reduced for all energies, and one can expect that as the amount of disorder increases, the total transmission will be dominated by small $k_\bot$ contributions. We believe that these findings would provide a strong motivation for study of Sb surface with steps 
along $\bar{\Gamma}-\bar{K}$ direction.

{\it Summary.---}In conclusion, we have performed an {\it ab-initio} study of the topological
surface states on the Sb(111) surface and their response to the presence of
single bilayer high steps, showing excellent agreement with experimental observations.  We have identified the various scattering
processes possible and formulated general conditions that lead to the
formation of the dominant scattering features. This enabled us to confirm the
fascinating chiral spin texture of the surface states.  Resonant tunneling
transmission across surface barriers, indicative of the extended nature of
these states, was found. We identified phase shifts in the scattered PDOS at quantum
well states energies and evaluated their life-times. The results demonstrate that it is possible to fully characterize the scattering properties of the barriers with first principles calculations. Finally, we have shown
that one can have minimal scattering along other high-symmetry
directions even in presence of strong surface perturbations, which provides a
unique signature for the topologically protected nature of these states. We
believe that this can be readily tested in future experiments.

{\it Acknowledgments.---}This work is financially supported by the Irish Research Council for Science, Engineering and Technology (IRCSET) 
under the EMBARK initiative. Computational resources have been provided by the Trinity Centre for High 
Performance Computing (TCHPC). I.R. and S.S. acknowledge additional financial support by KAUST (ACRAB project).
The authors would like to thank Anna Pertsova for illuminating discussions.

\end{document}